\documentclass{article}
 
\usepackage[english]{babel}

\usepackage[letterpaper,top=2cm,bottom=2cm,left=3cm,right=3cm,marginparwidth=1.75cm]{geometry}

\usepackage{amsmath,subfigure}
\usepackage{graphicx}
\usepackage{xcolor}
\usepackage[colorlinks=true, allcolors=blue]{hyperref}

\title{GenoCraft: A Comprehensive, User-Friendly Web-Based Platform for High-Throughput Omics Data Analysis and Visualization}

\author{Yingzhou Lu$^{1,2}$, Minjie Shen$^{2}$, Ling Yue$^{3}$, Chenhao Li$^{4}$, Lulu Chen$^{2}$, Xiao Wang$^{5}$, \\ 
David Herrington$^{6}$, Yue Wang$^2$, Yue Zhao$^{7}$, Tianfan Fu$^{3}$, Capucine Van Rechem$^{1,\ddag}$ 
}

\begin{document}
\maketitle

\begin{center}
$^{1}$Department of Pathology, Stanford University School of Medicine, Stanford, CA, 94305, USA  \\ 
$^{2}$ Bradley Department of Electrical and Computer Engineering, Virginia Tech, Blacksburg, VA, 24060, USA\\
$^{3}$Department of Computer Science, Rensselaer Polytechnic Institute, Troy, NY, 12180, USA \\
$^{4}$ Computer Engineering, University of Illinois Urbana-Champaign, Urbana, IL, 61820, USA\\  
$^{5}$Department of Computer Science and Engineering, University of Washington, 1410 NE Campus Parkway Seattle, WA 98195, USA\\
$^{6}$Department of Internal Medicine, Wake Forest University, Winston-Salem, NC, 27109, USA\\
$^{7}$ Department of Computer Science, University of Southern California, Los Angeles, CA, 90089, USA \\  
\ddag Lead contact: cvrechem@stanford.edu\\
\end{center}

\section*{Summary}
The surge in high-throughput omics data has reshaped the landscape of biological research, underlining the need for powerful, user-friendly data analysis and interpretation tools. 
This paper presents GenoCraft, a web-based comprehensive software solution designed to handle the entire pipeline of omics data processing. GenoCraft offers a unified platform featuring advanced bioinformatics tools, covering all aspects of omics data analysis. It encompasses a range of functionalities, such as normalization, quality control, differential analysis, network analysis, pathway analysis, and diverse visualization techniques.
This software makes state-of-the-art omics data analysis more accessible to a wider range of users.
With GenoCraft, researchers and data scientists have access to an array of cutting-edge bioinformatics tools under a user-friendly interface, making it a valuable resource for managing and analyzing large-scale omics data.
The API with an interactive web interface is publicly available at 
\url{https://genocraft.stanford.edu/}. 
We also release all the codes in \url{https://github.com/futianfan/GenoCraft}. 

\section{Introduction}
During the past few years, the field of genomics has seen a remarkable and unprecedented expansion, fueled by an enormous volume of data. High-throughput sequencing platforms have provided unprecedented opportunities to delve into the genomic landscape, heralding a surge in omics data, including genomics, transcriptomics, proteomics, and metabolomics~\cite{romero2006use,huang2022artificial}. However, while generating omics data has become increasingly accessible, analyzing these complex datasets remains a significant challenge to researchers and practitioners.

The handling and interpretation of the vast amounts of data generated by high-throughput technologies necessitate sophisticated computational tools~\cite{chen2016reads}. 
An intricate analysis pipeline is required to gain meaningful insights into these data, including quality control, preprocessing, alignment, variant calling, differential expression analysis, pathway analysis, etc~\cite{subramanian2005gene}. 
In order to carry out each of these steps, researchers need a specialized bioinformatics tool. Many researchers find these tools difficult to learn, as they require substantial computational resources~\cite{hwang2018single}.

GenoCraft aims to fill this gap by providing a comprehensive, intuitive platform that facilitates seamless navigation through complex omics data analysis. This web-based software solution integrates a suite of state-of-the-art bioinformatics tools within a user-friendly interface, thereby democratizing access to advanced omics data analysis. GenoCraft is not merely a tool; it is a versatile companion designed to empower researchers, enabling them to harness the full potential of omics data in driving breakthroughs in their respective fields~\cite{ilicic2016classification}.

GenoCraft contributes significantly to the field of omics data analysis in several key ways:
\begin{enumerate}
\item \textbf{Comprehensive Analytical Tools}: The GenoCraft platform integrates state-of-the-art bioinformatics tools across the entire omics data analysis pipeline, including normalization, quality control, differential analysis, network analysis, and pathway analysis. This comprehensive suite of tools provides a one-stop solution for researchers, eliminating the need to switch between different software or platforms for different stages of analysis.
\item \textbf{User-Friendly Interface}: Despite the complex nature of the analyses it performs, GenoCraft's user-friendly interface makes advanced bioinformatics accessible to researchers regardless of their programming background.
\item \textbf{High-Quality Visualization}:  The ability to effectively visualize complex data is a key component in turning raw data into meaningful biological insights. GenoCraft provides a wide range of data visualization options, facilitating the easy and intuitive interpretation of analysis results. 
\end{enumerate}

\section{Methods}

\begin{figure}
 \centering
 \includegraphics[width=0.8\textwidth]{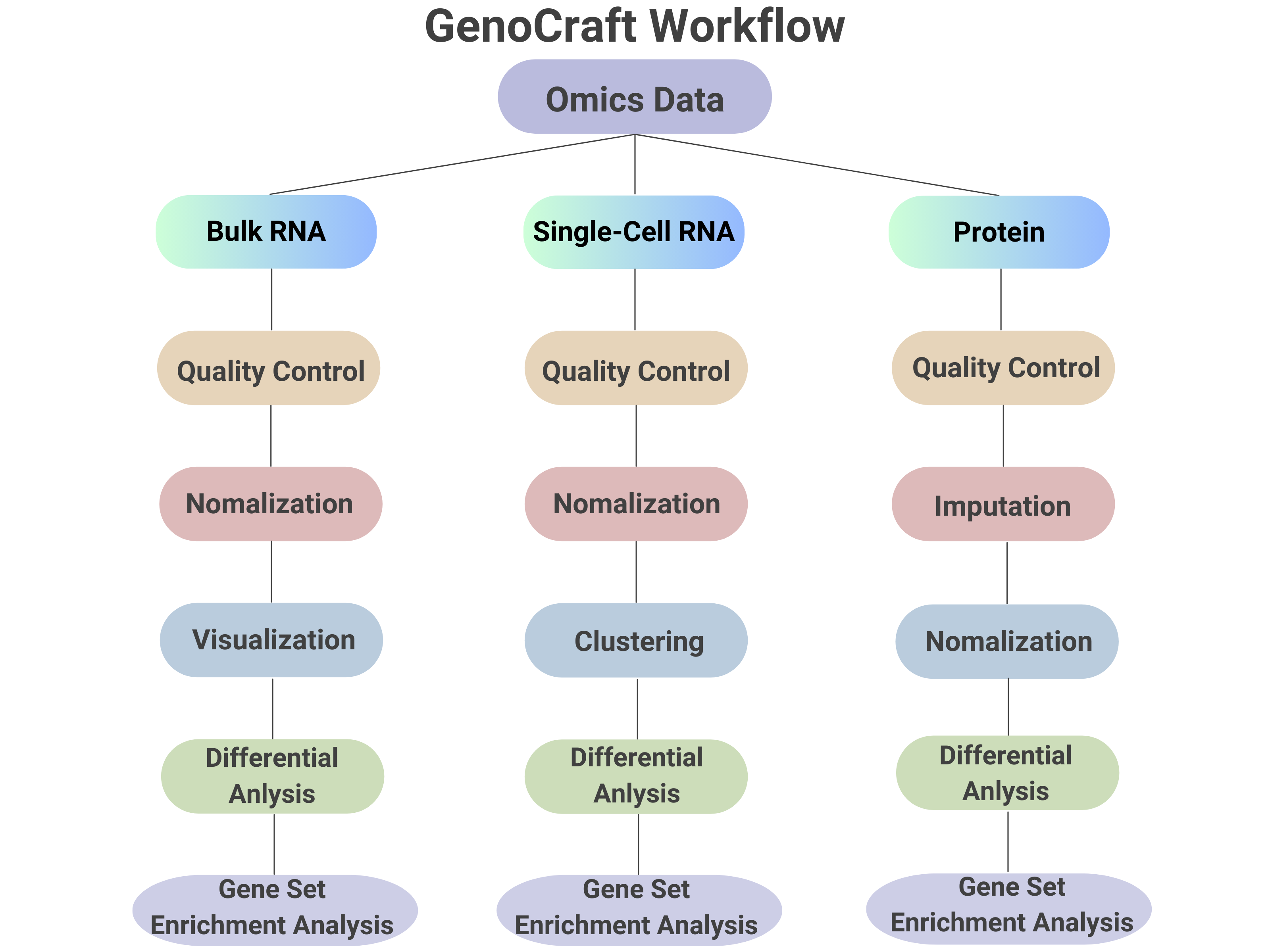}
 \caption{ Pipeline. }
 \vspace{-0.1in}
 \label{fig:pipeline}
\end{figure}

The following steps are the methodology employed in GenoCraft's data analysis pipeline~\cite{hwang2018single}. 
This pipeline incorporates a series of critical analytical stages, including normalization, quality control, differential analysis, network analysis, and pathway analysis, each contributing to a holistic understanding of complex omics data~\cite{holt2011maker2}. These components ensure a robust and in-depth exploration of the dataset, enabling the elucidation of meaningful and actionable biological insights.
For ease of exposition, we show the pipeline in Figure~\ref{fig:pipeline}.

\subsection{Quality Control}
Quality control is a crucial step in any data analysis pipeline \cite{lease2011quality}, especially in genomics and other omics studies~\cite{wang2012rseqc}. 
The reasons for performing quality control include:

\begin{itemize}
\item Identifying errors and outliers: Quality control helps detect any technical errors that might have occurred during the data collection process, such as issues with sample handling, sequencing, or data extraction~\cite{wang2012rseqc}. It also helps identify any outliers that may skew the analysis results.

\item Ensuring data integrity: By checking the data quality, researchers can ensure that the data used in their analysis is accurate and reliable. This step is critical to validate the conclusions drawn from the analysis~\cite{yamada2021interpretation}.

\item Correcting systematic biases: Quality control can help identify and correct systematic biases in the data, such as batch effects or sample contamination, which can influence the results.
\end{itemize}
The quality control module in GenoCraft performs a comprehensive evaluation of data quality to detect potential issues early in the analysis process. The quality control method implemented in this script is aimed at refining a dataset of gene counts. This is achieved by excluding genes that do not meet the predefined threshold of minimum counts (default value 10) and minimum number of samples (default value 3). The minimum counts specify how many times a gene must appear in the dataset, while the minimum samples specify the number of samples in which the gene must appear. This filtering process reduces the size of the dataset by removing low-frequency genes, thus enhancing the reliability of any subsequent data analysis.

\subsection{Normalization}

Normalization is a preliminary step in the analysis of high-throughput omics data~\cite{ding2015normalization}. GenoCraft employs a sophisticated algorithm for normalization to correct for technical variations such as different sequencing depths or RNA composition, ensuring comparability across samples~\cite{ding2015normalization,wu2022cosbin}. The specific normalization technique used is dependent on the nature of the data, with options including Total Count, RPKM (reads per kilobase of transcript per million reads mapped)/FPKM (fragments per kilobase of transcript per million fragments mapped), TPM (transcripts per million), and others for RNA-seq, or TMM (Trimmed Mean of M-values) for gene expression~\cite{du2023abds,zhao2020misuse}. 

In the CPM (Counts Per Million) normalization procedure, the raw read counts for each gene in a sample are transformed into proportions relative to the total number of mapped reads per sample. This transformation allows for comparisons of gene expression levels across different samples. The CPM normalization method follows the approach described by Robinson et al. in~\cite{anders2013count}.

The counts per gene were normalized to CPM by dividing each count by the total number of mapped reads in the sample and multiplying by $1 \times 10^6$. This normalization method enables the comparison of gene expression levels across samples.

\subsection{Differential Analysis}
Identifying markers in genomics and other omics studies is of paramount importance as these markers often represent genes, proteins, or other molecules that exhibit significant differences between conditions, such as healthy and diseased subtypes~\cite{chen2021data}. 

Differential analysis is a core component of GenoCraft's analytical repertoire. The software utilizes robust statistical models to identify genes, transcripts, or proteins that exhibit significant differences in expression between conditions~\cite{lu2022cot}. 

The t-test, integral to the process of marker identification in GenoCraft, is leveraged to compare the mean values between two distinct groups to ascertain their statistical divergence~\cite{rouder2009bayesian}. Within the GenoCraft system, a predefined threshold is established to filter and highlight genes that surpass this level, demonstrating statistically significant differences, thereby ensuring robust and reliable identification of key markers.

\subsection{Network Analysis}
Complex diseases frequently exhibit the mismanagement of several critical biological pathways. Differential network analysis, a tool that seeks to identify the reorganization of regulatory correlation structures under varying biological conditions, is crucial for understanding the molecular roots of disease progression and treatment reactions.

GenoCraft incorporates differential network analysis, known as Differential Dependency Networks (DDN3.0)~\cite{zhang2021ddn2}, which is skilled at concurrently learning both common and rewired network structures.
Network analysis identifies in a complex and often unknown overall molecular circuitry a network of differentially connected molecular entities.
Researchers can investigate these networks to uncover modules of interconnected genes or proteins, identify central or ``hub'' nodes, and explore how signals propagate through the network~\cite{marbach2012wisdom}, and further help to provide a plausible interpretation of data, gain new insight of disease biology, and generate novel hypotheses for further validation and investigations.

\subsection{Gene Set Enrichment Analysis}

Gene Set Enrichment Analysis (GSEA), also called pathway analysis, represents a computational technique that assesses whether a pre-established collection of genes exhibits statistically significant and consistent differences when comparing two distinct biological conditions. (e.g., phenotypes). 

GenoCraft's pathway analysis tools allow users to move beyond lists of differentially expressed genes to uncover the biological processes that are enriched in their data. 

By mapping genes onto known pathways from databases such as KEGG (Kyoto Encyclopedia of Genes and Genomes)~\cite{kanehisa2000kegg} or Reactome~\cite{fabregat2018reactome}, GenoCraft provides insights into the biological mechanisms underlying the observed changes in gene or protein expression~\cite{subramanian2005gene}. The software also employs statistical methods to identify pathways that are significantly enriched (with p-values provided), providing a clear direction for further investigation~\cite{kuleshov2016enrichr}.

These comprehensive and integrated analyses performed by GenoCraft ensure a thorough exploration and understanding of the dataset in hand, enabling researchers to glean meaningful biological insights from their omics data~\cite{lu2018multi}.

\subsection{Clustering}

A clustering algorithm is a machine learning technique used to group similar data points together based on their inherent similarities or patterns~\cite{hartigan1979algorithm}. It aims to partition a dataset into subsets or clusters, where data points within each cluster are more similar to each other than to those in other clusters~\cite{aibar2017scenic}. In the context of omics data, clustering techniques categorize genes, proteins, or samples into groups, such that entities within a cluster show higher similarity to each other based on the chosen metric and less similarity to entities in other clusters~\cite{pavlopoulos2011using}.

GenoCraft supports various clustering methods, including hierarchical clustering and k-means clustering.
The clustering results in GenoCraft can be visualized using heatmaps, dendrograms, or multi-dimensional scaling plots, providing intuitive graphical representations of the patterns in the data.

\subsection{Visualization}

t-Distributed Stochastic Neighbor Embedding (t-SNE) is a non-linear dimensionality reduction technique specifically designed for high-dimensional data, often employed to visualize complex data structures. Developed by Laurens van der Maaten and Geoffrey Hinton, it stands out due to its ability to preserve local structures within the data. The method ensures that data points that are close to each other in the high-dimensional dataset remain close when projected onto a lower-dimensional space~\cite{van2008visualizing}. 
t-SNE works by converting the high-dimensional Euclidean distances between data points into conditional probabilities that represent similarities. 
Data points that are close in the high-dimensional space have a higher probability of being picked, whereas data points that are distant have a lower probability. This conditional probability is then translated into a two or three-dimensional map, creating a 'neighborhood' of points where the structure of the neighborhood attempts to reflect the structure in the original high-dimensional space as closely as possible. By doing this, t-SNE provides a way to visually explore patterns, clusters and outliers in high-dimensional data, making it a very useful tool in exploratory data analysis.

\section{Results}

In this section, we demonstrate some case studies to illustrate how to use GenoCraft. 
The API is publicly available at \url{https://genocraft.stanford.edu/}. 
An interactive web interface is available. 
We also release all the codes in \url{https://github.com/futianfan/GenoCraft}. 

\subsection{Case Study: GSE152418}
In this case study, we applied GenoCraft’s RNA-seq analysis pipeline to the GSE152418 dataset from the Gene Expression Omnibus (GEO). This dataset explores the immune response in patients with COVID-19 by analyzing gene expression in peripheral blood mononuclear cells (PBMCs) from both COVID-19 patients and healthy convalescent controls.

The raw counts for gene expression were retrieved from GSE152418, which consists of samples from COVID-19 patients and convalescent controls. The dataset was structured with gene identifiers (ENSEMBL IDs) as rows and samples as columns, including 17 COVID-19 patients and their respective control counterparts. Initial quality control was performed by removing samples and genes with low counts, ensuring the data was suitable for downstream analysis.

To account for differences in sequencing depth and library sizes across samples, we applied Counts Per Million (CPM) normalization to the raw read counts. This normalization technique adjusts the gene expression values, making them comparable across the dataset. CPM is particularly useful when working with RNA-seq data, as it mitigates the effects of varying sequencing depths across samples.

We proceeded with the differential expression analysis, comparing the COVID-19 group against the convalescent control group. Using GenoCraft’s integrated t-test, we calculated log fold changes (LFC) between the two groups to identify genes that were significantly upregulated or downregulated in COVID-19 patients.

Significant genes were identified based on their fold change and statistical significance. These genes represent key players in the immune response to COVID-19 and provide valuable insights into the molecular mechanisms driving disease pathology.

The differentially expressed genes (DEGs) were further subjected to pathway enrichment analysis using the well-known databases. By mapping DEGs to known immune response pathways, we identified critical pathways activated in COVID-19 patients. These findings revealed potential biomarkers and therapeutic targets relevant to the immune dysregulation observed in COVID-19.

GenoCraft’s suite of visualization tools provided intuitive representations of the analysis results. Heatmaps illustrated the expression patterns of the most significant genes across samples, while volcano plots highlighted genes with the highest fold changes. These visualizations offered a clear and comprehensive overview of the data, facilitating interpretation and aiding in the identification of biologically relevant gene expression changes.

The analysis of the GSE152418 dataset using GenoCraft underscores the platform’s capabilities in handling large-scale RNA-seq datasets. By identifying differentially expressed genes and key immune pathways, GenoCraft provided actionable insights into the biological response to COVID-19, potentially informing future research and therapeutic development.

\begin{figure}[ht]
\centering
\includegraphics[width=0.7\textwidth]{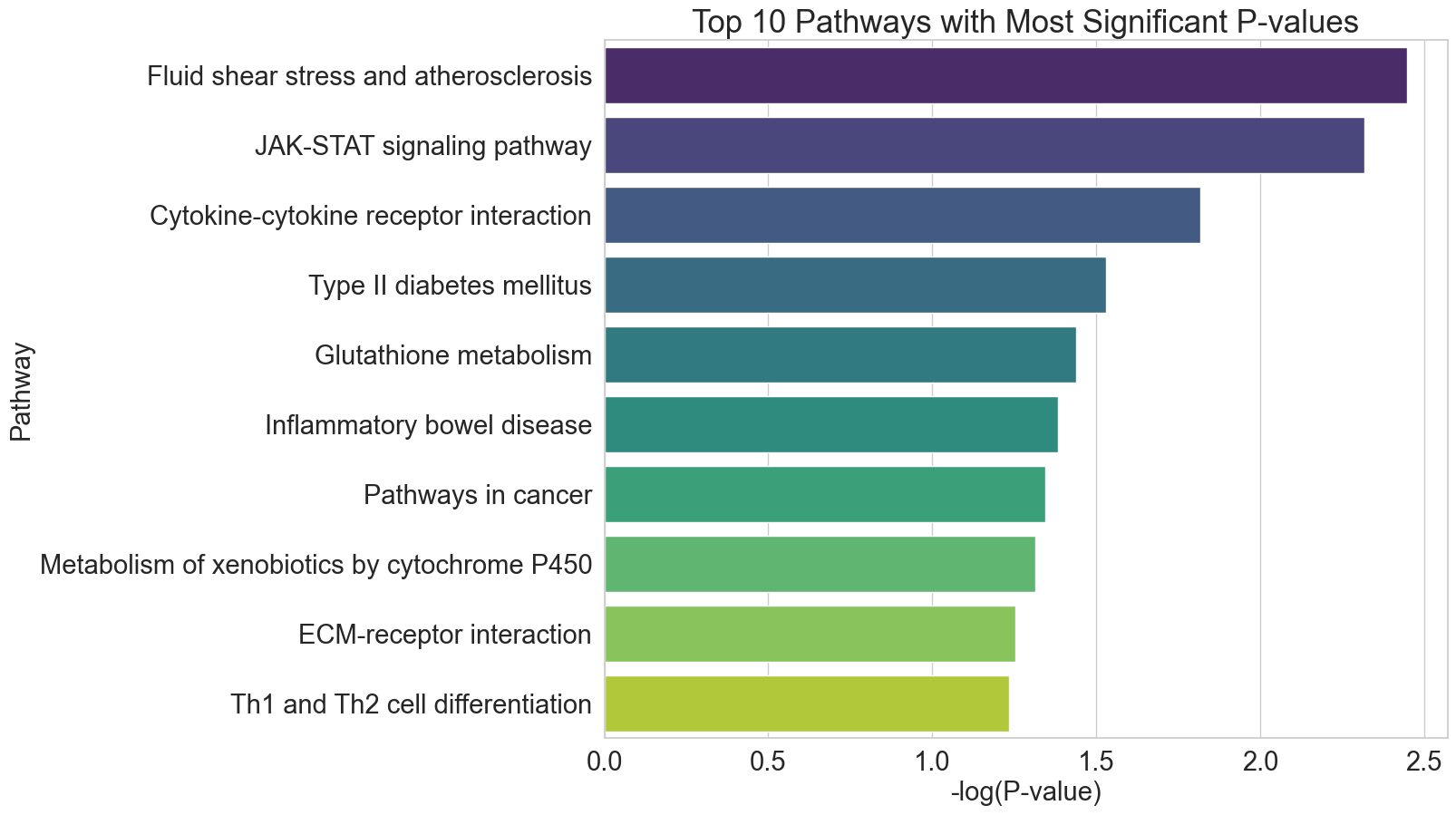}
\vspace{-0.1in}
\caption{bulk RNA pathway analysis}
\label{fig:bulk_GSEA.png}
\end{figure}

By going through the stages of data retrieval, quality check, gene expression quantification, and differential expression analysis, we aim to pinpoint genes whose expression levels significantly differ across conditions. This can provide valuable information on the biological conditions and phenomena under study, potentially contributing to advancements in fields such as disease research, drug development, and understanding of biological processes. This case study also serves to demonstrate the practical applications of various bioinformatics tools and packages in a typical RNA-seq analysis pipeline, highlighting the importance of computational methods in modern biological research.

\subsection{Case Study 2: Single-Cell Analysis of GSE69405}

The dataset GSE69405 (\url{https://ftp.ncbi.nlm.nih.gov/geo/series/GSE69nnn/GSE69405/matrix/}) presents a rich resource for understanding the intricacies of lung adenocarcinoma at the single-cell level. 
This dataset leverages high-throughput sequencing to profile the transcriptomes of single cancer cells derived from patient lung adenocarcinoma xenografts (PDX) tumors~\cite{hahn1995allelotype}. The overarching aim was to investigate the impact of intratumoral heterogeneity on anti-cancer drug responses.

\begin{figure}[ht]
\centering
\includegraphics[width=0.60\textwidth]{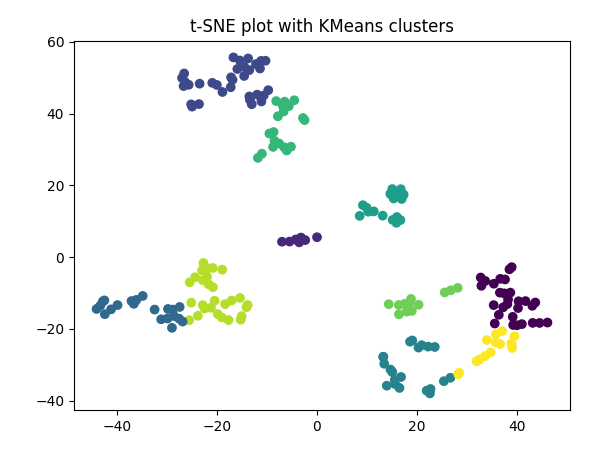}
\vspace{-0.1in}
\caption{Single-cell t-SNE. }
\vspace{-0.1in}
\label{fig:single-cell-tsne}
\end{figure}

For the pipeline analysis, we first performed quality control and normalized the raw counts from the bulk RNA sequencing data. After that, we utilized hierarchical clustering (Figure~\ref{fig:single-cell-tsne}) to group samples based on their gene expression profiles. We then performed gene-level differential expression analysis to identify genes with significant changes in expression between the different conditions. This provided insights into the similarities and differences between the samples and allowed us to identify potential biomarkers and therapeutic targets for lung adenocarcinoma.

\begin{figure}[ht]
    \centering
    \includegraphics[width=0.9\textwidth]{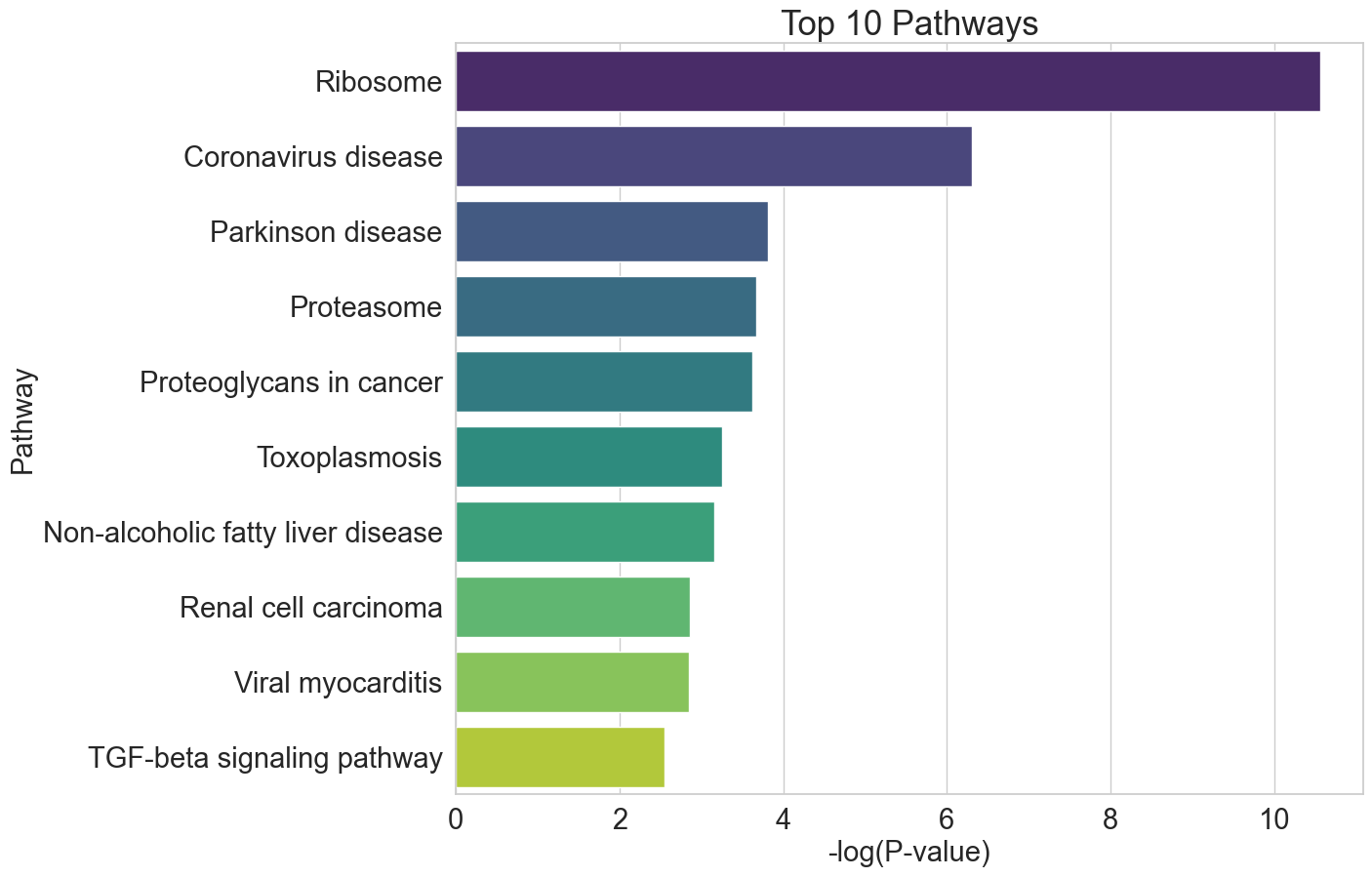}
    \caption{Single-Cell pathway analysis. }
    \label{fig:single-cell pathway analysis}
\end{figure}

Subsequent pathway analysis (Figure~\ref{fig:single-cell pathway analysis}), incorporating the identified differentially expressed genes, unveiled the biological processes and pathways that are most significantly altered in the cancer cells. Our findings underscored the complex nature of intratumoral heterogeneity and its potential influence on the effectiveness of anti-cancer drugs.

In conclusion, the comprehensive analysis of the GSE69405 dataset provides valuable insights into the heterogeneity of lung adenocarcinoma at the single-cell level and its implications for drug responses. This case study demonstrates the power of combining single-cell and bulk RNA sequencing methodologies to unravel the complex landscape of cancer biology.

\subsection{Case Study 3: Protein}

Rhabdoid tumors, highly aggressive childhood cancers, are driven by the loss of the mSWI/SNF (mammalian SWItch/Sucrose Non-Fermentable) subunit SMARCB1. These tumors, which can develop in the brain, kidneys, or soft tissues, exhibit specific sensitivity to the translation inhibitor homoharringtonin (HHT). 

In this computational biology-driven analysis, we utilized GenoCraft to perform a comprehensive examination of protein expression data from rhabdoid tumor cells, focusing on identifying significant proteins and their associated pathways. The protein dataset was first subjected to quality control and normalization processes to ensure the reliability of the results. 

The differential expression analysis revealed a set of 24 proteins with statistically significant changes between the conditions under study. These proteins were identified based on an adjusted p-value threshold of 0.02 and a log2 fold change (LogFC) greater than 1.5 for upregulated proteins. The volcano plot generated from this analysis provided a clear visual representation of the significant proteins, with those meeting the criteria for differential expression highlighted in red. Notably, proteins associated with chromatin remodeling, particularly those within the SWI/SNF complex, emerged as key players in the dataset, suggesting their involvement in critical cellular processes.

Further pathway enrichment analysis provided additional insights into the biological functions of the significant proteins. The top 20 pathways, ranked by -log10(Adjusted p-value), included the SWI/SNF complex, ATPase complex, and pathways related to chromatin remodeling. These findings underscore the potential role of these pathways in driving the observed protein expression changes. By integrating differential expression analysis with pathway enrichment from GO, this case study highlights the utility of GenoCraft in uncovering biologically relevant insights that can inform future research directions and therapeutic development.

The results of this analysis not only enhance our understanding of the molecular mechanisms underlying the studied conditions but also demonstrate the effectiveness of GenoCraft as a comprehensive tool for high-throughput omics data analysis. The identification of key proteins and pathways provides a strong foundation for further exploration, potentially leading to the discovery of novel biomarkers or therapeutic targets.

\begin{figure}[ht]
\centering
\includegraphics[width=0.9\textwidth]{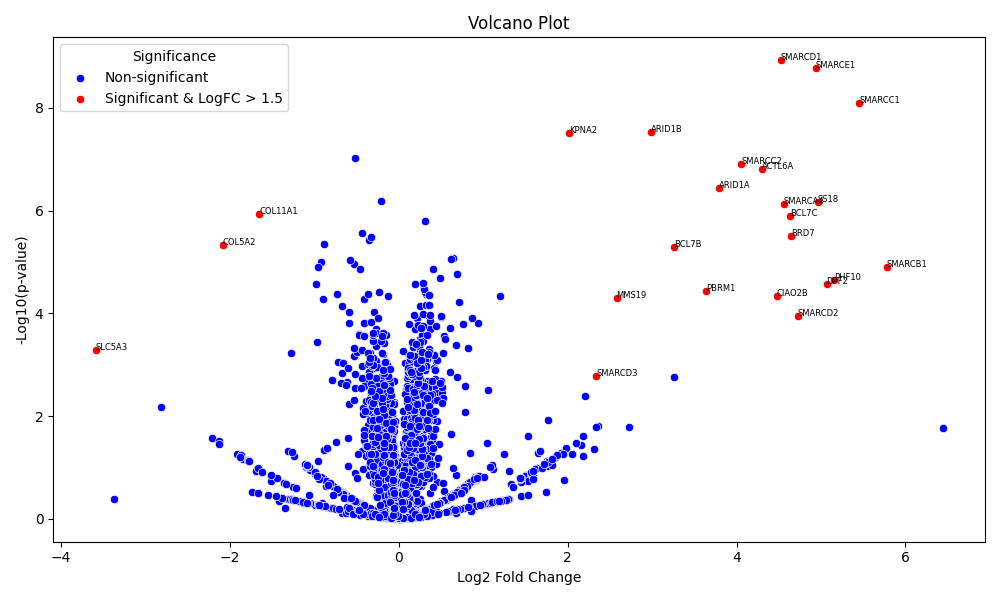}
\caption{significant protein markers }
\label{fig:volcano_plot_annotated}
\end{figure}

\begin{figure}[ht]
\centering
\includegraphics[width=0.6\textwidth]{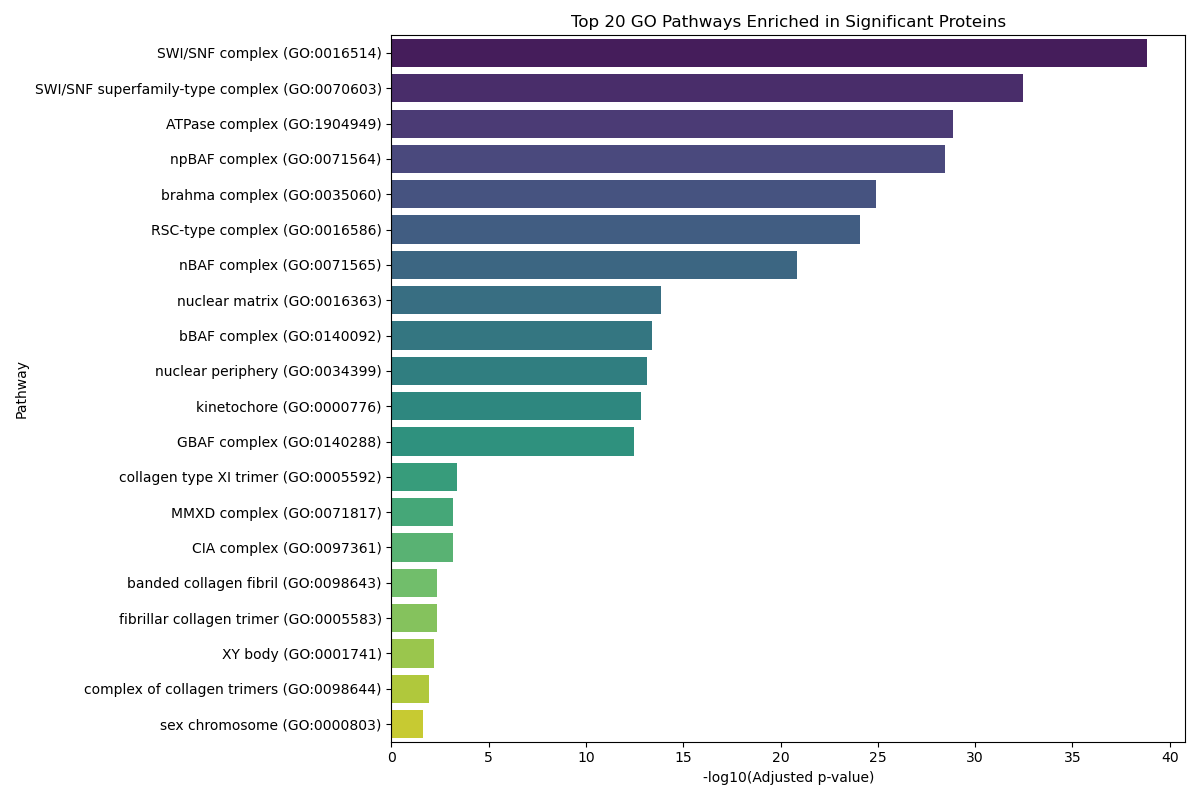}
\caption{Top 20 GO Pathways }
\label{fig:Protein_GO_Pathways_Top20}
\end{figure}

\section{Performance}
\subsection{Performance and Scalability}

In the dynamic realm of genomics, where data volumes are burgeoning and the complexity of analyses is escalating, the ability to process data swiftly and at a vast scale becomes paramount. It's not just about crunching numbers; it's about ensuring timely insights, especially when these insights could lead to groundbreaking discoveries or critical clinical decisions. Genocraft is conceived and engineered precisely to cater to these pressing needs.

From its inception, Genocraft was envisioned as more than just another tool in the genomic toolbox. It was designed as a comprehensive solution, capable of adapting to the multifaceted challenges posed by modern genomic research. This adaptability is not merely theoretical. Our benchmarks, highlighted in the \textbf{Results} section, provide empirical evidence of Genocraft's prowess. These metrics aren't just numbers; they signify the software's capability to deliver accurate results in a time-efficient manner, even when subjected to massive datasets or intricate analytical requirements.

Beyond the raw performance, it's essential to recognize the underlying factors that contribute to Genocraft's efficiency. Its algorithms, honed through rigorous testing and iterative refinement, are optimized for speed without compromising on accuracy. The data handling processes, streamlined through state-of-the-art computational techniques, ensure that Genocraft can ingest, process, and output data with minimal bottlenecks.

Scalability is another standout feature of GenoCraft, making it exceptionally suitable for the ever-expanding demands of genomic research. The tool is adept at efficiently processing large-scale omics datasets, a capability that is becoming increasingly crucial as the volume and complexity of genomic data continue to grow. GenoCraft's robust architecture and optimized algorithms allow for seamless scaling, accommodating datasets of varying sizes without compromising on processing speed or analytical accuracy. This scalability extends not only to data volume but also to the diversity of data types and computational complexity, ensuring that GenoCraft remains a versatile and powerful tool for genomics analysis.

Unlike peer methods that are often limited to specific steps, GenoCraft excels in executing a comprehensive range of tasks, from quality control to differential analysis and data visualization. Its unique ability to integrate various functionalities, including normalization, network analysis, and gene set enrichment analysis, within a user-friendly interface, distinguishes it from other existing tools. Due to this wide-ranging, end-to-end analytical capability, it is not practical to compare GenoCraft directly with other tools that are specialized in only certain aspects of the analytical process. This all-encompassing approach not only ensures a thorough and holistic data interpretation but also positions GenoCraft as a pioneering solution in bioinformatics, offering unprecedented efficiency and depth in omics data analysis.


\subsection{Computational Complexity}

In this paper, we introduce an innovative bioinformatics software tool tailored for comprehensive genomic analysis across the entire pipeline. Each module of the software has been developed with computational efficiency. The outlier detection algorithm, utilizing standard Z-score computations, operates with a linear complexity of O(N), ensuring swift processing even with large datasets.

For protein data, the missing data imputation, a critical step in data preprocessing, is executed using a KNNImputer, which inherently has a higher computational demand, scaling as $O(M * N * \log(N))$, where $M$ represents the number of features and $N$ the number of samples. This allows for accurate imputation but requires consideration of resource allocation for larger datasets.

For single-cell RNA seq data, the Principal Component Analysis (PCA) is the most computationally demanding step, with a complexity of $O(min(N^2M, NM^2))$, given the eigenvalue decomposition involved, particularly significant in scenarios where the number of genes surpasses the sample size. Following PCA, the K-Means clustering algorithm introduces a complexity of $O(I * K * NM)$, with $I$ being the iteration count, $K$ the number of clusters, and $N$ and $M$ the dimensions of the PCA-reduced dataset. 
The biomarker selection process, involving t-tests and feature sorting, primarily follows an $O(N)$ complexity for computations, with an additional $O(M\log M)$ for sorting, balancing computational efficiency with statistical rigor. Lastly, the downstream analysis module, integrating external API, adds a layer of complexity that is primarily dependent on external server response times.

Collectively, the software is optimized to handle large-scale genomics data, providing a balance between computational efficiency and analytical depth.

\section{Discussion}
GenoCraft, a state-of-the-art web-based platform, stands as a formidable powerhouse for high-throughput omics data analysis and visualization. By synergizing a diverse array of cutting-edge bioinformatics techniques, it has anchored its position as an indispensable resource for researchers delving deep into the intricate maze of omics data. Right from the initial steps of normalization and quality assurance to the more intricate facets of differential, network, and pathway analyses, GenoCraft delivers an exhaustive and meticulous data scrutiny~\cite{lu2019integrated,wang2023scientific}.

Adding to its appeal is its intuitive user interface complemented by a rich array of data visualization tools, optimizing both its accessibility and usability. This seamless integration facilitates researchers in swift and coherent data interpretation, bridging the gap between data collation and profound biological discoveries, thereby propelling advancements in a myriad of scientific domains~\cite{lu2023machine}.

In conclusion, GenoCraft's development has democratized access to advanced omics data analysis, lowering barriers to entry for researchers across the spectrum of expertise levels. This comprehensive platform continues to evolve in response to the rapidly advancing field of genomics, remaining committed to empowering researchers to unravel the complex mechanisms underlying biological systems.

As GenoCraft continues to evolve, its future directions are particularly promising in the realms of spatial genomics and multi-omics integration. The incorporation of spatial genomics will enable GenoCraft to map and analyze genomic data in the context of the physical locations within a cell or a tissue, thereby providing a more comprehensive understanding of cellular functions and disease mechanisms. This spatial resolution is expected to revolutionize the way genomic data is interpreted, offering deeper insights into the spatial heterogeneity of gene expression. Additionally, the integration of multi-omics data analysis is another exciting frontier for GenoCraft. By combining genomic, transcriptomic, proteomic, and metabolomic data, GenoCraft aims to offer a holistic view of the biological systems, facilitating a more thorough understanding of complex biological interactions and pathways. This multi-omics approach will not only enhance the depth of genomic analysis but also pave the way for more personalized and precise medical interventions. These future advancements position GenoCraft at the cutting edge of genomic research, ready to tackle the next generation of challenges in the field.

\section{Limitation of the study}
This study is currently limited to RNA and proteomics data. While this focus has allowed for a detailed exploration within this domain, it inherently narrows the scope of our findings, as they may not be directly applicable to other omics types. Recognizing the interconnected nature of various genomic disciplines, we acknowledge that our results represent a piece of the broader genomic puzzle. In future research, we plan to extend our analysis to include additional omics types, such as genomics, transcriptomics, and metabolomics. This expansion will not only enhance the comprehensiveness of our findings but also provide a more holistic view of genomic interactions and functions.

\section{Acknowledgements}
Work related to this study is supported by the Sontag Foundation (C.V.R), the American Cancer Society (C.V.R.) and the Department of Defense Breast Cancer Research Program (C.V.R.).

\section{Contributions}
Y.L., T.F, and Y.W. developed the Genocraft method and algorithm. T.F. and F.M. applied the Genocraft framework and performed data analyses. Y.L. and C.V.R. interpreted the data and wrote the manuscript. M.S., L.Y. and T.F. designed the user interface and the front-end and back-end implementation. D.H and C.V.R. supervised the study.

\section{Conflicts of Interest}
All authors declare that they have no conflicts of interest.

\section{Declaration of generative AI and AI-assisted technologies in the writing process}

During the preparation of this work, we used chatGPT and Wordtune in order to correct the grammar and improve the language. After using this tool/service, the authors reviewed and edited the content as needed and take full responsibility for the content of the publication.

\bibliographystyle{alpha}
\bibliography{sample}

\end{document}